\documentclass[aps,prl,twocolumn,superscriptaddress,showpacs]{revtex4}
\usepackage{graphicx}
\usepackage{latexsym}
\usepackage{amssymb}
\usepackage{amsmath}
\usepackage{amsfonts}
\usepackage{bm}
\usepackage{multirow}
\usepackage{color}
\usepackage{comment}
\newcommand{\ii}{\mathrm{i}}

\newcommand{\SO}{\mathrm{SO}}
\renewcommand{\O}{\mathrm{O}}
\newcommand{\SU}{\mathrm{SU}}

\newcommand{\beq}{\begin{equation}}
\newcommand{\eeq}{\end{equation}}
\newcommand{\beqn}{\begin{eqnarray}}
\newcommand{\eeqn}{\end{eqnarray}}

\newcommand{\nop}[2]{n^{#2}(#1,\bar{#1})}
\newcommand{\vop}[1]{v(#1,\bar{#1})}

\DeclareMathAlphabet{\mathbbold}{U}{bbold}{m}{n}

\def\SU{{\rm SU}}

\begin{document}

\title{Continuous N\'{e}el-VBS Quantum Phase Transition in Non-Local one-dimensional systems with SO(3) Symmetry}

\author{Chao-Ming Jian}
\affiliation{Station Q, Microsoft, Santa Barbara, California
93106-6105, USA}

\author{Yichen Xu}
\affiliation{Department of Physics, University of California,
Santa Barbara, CA 93106, USA}

\author{Xiao-Chuan Wu}
\affiliation{Department of Physics, University of California,
Santa Barbara, CA 93106, USA}

\author{Cenke Xu}
\affiliation{Department of Physics, University of California,
Santa Barbara, CA 93106, USA}

\begin{abstract}

One dimensional $(1d)$ interacting systems with local Hamiltonians
can be studied with various well-developed analytical methods.
Recently novel $1d$ physics was found numerically in systems with
either spatially nonlocal interactions, or at the $1d$ boundary of
$2d$ quantum critical points, and the critical fluctuation in the
bulk also yields effective nonlocal interactions at the boundary.
This work studies the edge states at the $1d$ boundary of $2d$
strongly interacting symmetry protected topological (SPT) states,
when the bulk is driven to a disorder-order phase transition. We
will take the $2d$ Affleck-Kennedy-Lieb-Tasaki (AKLT) state as an
example, which is a SPT state protected by the $\SO(3)$ spin
symmetry and spatial translation. We found that the original
$(1+1)d$ boundary conformal field theory of the AKLT state is
unstable due to coupling to the boundary avatar of the bulk
quantum critical fluctuations. When the bulk is fixed at the
quantum critical point, within the accuracy of our expansion
method, we find that by tuning one parameter at the boundary,
there is a generic direct transition between the long range
antiferromagnetic N\'{e}el order and the valence bond solid (VBS)
order. This transition is very similar to the N\'{e}el-VBS
transition recently found in numerical simulation of a spin-1/2
chain with nonlocal spatial interactions. Connections between our
analytical studies and recent numerical results concerning the
edge states of the $2d$ AKLT-like state at a bulk quantum phase
transition will also be discussed.

\end{abstract}

\maketitle

%\section{Introduction}

Our understanding of one dimensional $(1d)$ quantum many-body
systems with local Hamiltonians is far more complete compared with
higher dimensional systems, since many powerful analytical methods
such as Bethe ansatz~\cite{bethe}, Virasoro algebra~\cite{cft1},
etc. are applicable only to $1d$ systems (or $(1+1)d$ space-time).
We also understand that $1d$ systems have many unique features
that are fundamentally different from higher dimensions. For
example, with local Hamiltonians, generally there can not be
spontaneous continuous symmetry breaking in $(1+1)d$ even at zero
temperature (with exceptions of the scenarios when a fully
polarized ferromagnet is the exact ground state), the closest one
can possibly get is a quasi-long range power-law correlation of
order parameters that transform nontrivially under a continuous
symmetry. There is also no topological order in $1d$ systems
analogous to fractional quantum Hall states which have a gap and
simultaneously ground state topological
degeneracy~\cite{wenreview2}. This means that many phenomena that
are found in higher dimensions do not occur in $1d$ systems.

To seek for richer physics in one dimensional systems, we need to
explore beyond the restriction of local Hamiltonians. One way to
get around this restriction is to consider $1d$ systems at the
boundary of a $2d$ systems, and drive the $2d$ bulk to a quantum
phase transition. The physics becomes especially interesting when
the disordered phase in the phase diagram of the $2d$ bulk is a
symmetry protected topological (SPT) phase, which already has
topologically protected $1d$ edge state. The interplay between the
topological edge state and gapless quantum critical modes can lead
to very nontrivial physics, which has been studied through
numerical methods recently~\cite{zhang1,zhang2,stefan1,stefan2}.
One can also directly turn on nonlocal spatial interaction in a
$1d$ Hamiltonian. $1d$ quantum spin chains with nonlocal spatial
interactions have also been studied recently, and very intriguing
physics was found~\cite{sandvik1,sandvik2}. We will discuss the
results of these numerical works later in this paper.

In this work we investigate the $2d$ SPT state protected by
symmetry $\SO(3) \times G$, where $\SO(3)$ is the ordinary spin
symmetry, while $G$ is a discrete symmetry, which could be an
onsite unitary $Z_2$ symmetry, or an anti-unitary time-reversal
$Z_2^T$. $G$ can also be a lattice symmetry such as translation by
one lattice constant. For example, when $G$ is the translation
along the $\hat{x}$ axis ($T_{x}$), this state can be realized as
the Affleck-Kennedy-Lieb-Tasaki (AKLT) state of the spin-2 system
on a $2d$ square lattice~\cite{AKLT}. In the example of spin-2
AKLT state, there is a chain of dangling spin-1/2 at the boundary
of the system, as long as the boundary is along the $\hat{x}$ axis
and preserves the translation symmetry $T_x$. The nature of the
SPT states, and the Lieb-Shultz-Mattis (LSM)
theorem~\cite{LSM,oshikawa,hastings} guarantee that this boundary
system cannot be trivially gapped, $i.e.$ it must be either
gapless, or gapped but degenerate (For a closed $1d$ system
without $0d$ boundaries, a generic ground state degeneracy can
only originate from spontaneous discrete symmetry
breaking~\cite{wenreview2}). In this work we will take the AKLT
state as an example, but our results can be straightforwardly
generalized to other discrete symmetries $G$.

Our study will mainly focus on the $1d$ boundary of strongly
interacting $2d$ bosonic SPT phases, using a controlled
renormalization group method. We would like to mention that
previous literature has discussed the coupling between quantum
criticality and topologically localized gapless states in various
fermionic topological insulators~\cite{groveredge}; other
approaches such as constructing soluble models and various
numerical methods have also been used to study edge states of
interacting SPT states at a bulk quantum
criticality~\cite{scaffidi2017,poll,poll2}. Our main finding is
that there is a generic continuous quantum phase transition
between a long range antiferromagnetic N\'{e}el order which
spontaneously breaks the $\SO(3)$ spin symmetry, and a valence
bond solid state, at the $1d$ boundary of an AKLT state that
couples to the bulk quantum critical modes. The bulk quantum
critical modes effectively yield nonlocal interactions at the $1d$
boundary, which makes the long range N\'{e}el order possible.

In principle the $1d$ boundary of this AKLT state should be
effectively described by an extended Heisenberg model \beqn H =
\sum_{j} J \vec{S}_j \cdot \vec{S}_{j+1} + \cdots \label{H}\eeqn
where $\vec{S}_j$ is the spin-1/2 operator, and the ellipsis
includes other possible terms allowed by $\SO(3) \times T_x$. The
ground state of Eq.~\ref{H} depends on the entire lattice
Hamiltonian. But a useful starting point of analyzing this
boundary system is the $\SU(2)_1$ conformal field theory (CFT)
described by the following Hamiltonian in the infrared limit:
\beqn H_0 = \int dx \ \frac{1}{3\cdot 2\pi} \left( \vec{J}_{L}
\cdot \vec{J}_L + \vec{J}_{R} \cdot \vec{J}_R \right). \label{H0}
\eeqn The $\SU(2)_1$ CFT has a larger symmetry than the lattice
Hamiltonian Eq.~\ref{H0}, since $\vec{J}_L$ and $\vec{J}_R$
generate the $\SU(2)_{L,R}$ symmetries for the left and right
chiral modes respectively. The relation between the microscopic
operator $\vec{S}$ and the low energy field is~\cite{balents2004}
\beqn \vec{S}(x) \sim \frac{1}{2\pi} \left(\vec{J}_L(x) +
\vec{J}_R(x) \right) + (-1)^x \vec{n}(x), \eeqn where $\vec{n}(x)$
is the N\'{e}el order parameter at the boundary. $\vec{J}_{L,R}$
both have scaling dimension $+1$ at the $\SU(2)_1$ CFT fixed
point, while $\vec{n}(x)$ has scaling dimension $1/2$ at the
$\SU(2)_1$ CFT.

The diagonal $\SU(2)$ symmetry (simultaneous $\SU(2)$ rotation
between the left and right modes) corresponds to the original
$\SO(3)$ spin symmetry on the lattice scale. And because the
lattice Hamiltonian has a lower symmetry than the infrared theory
Eq.~\ref{H0}, another term is allowed in the low energy
Hamiltonian: \beqn H_1 = \int dx \ \lambda \vec{J}_L \cdot
\vec{J}_R. \label{H1} \eeqn Since $\vec{J}_{L,R}$ have scaling
dimension $+1$, power-counting indicates the coefficient $\lambda$
has scaling dimension 0. Depending on the sign of $\lambda$, this
term can be either marginally relevant or marginally irrelevant.
When $\lambda$ is negative and marginally irrelevant the system
flows back to the $\SU(2)_1$ CFT with an enlarged $\SU(2)_L \times
\SU(2)_R$ symmetry. When this term is positive and marginally
relevant, it will flow to infinite (nonperturbative) and generate
a mass gap, which based on the nature of the SPT phase would imply
that the system spontaneously breaks the discrete symmetry $G$.
For example, when this system is realized as the AKLT state, and
$G$ is the translation $T_x$, the LSM theorem demands that when
the boundary of the system generates a mass gap, it spontaneously
breaks the translation symmetry and develops a nonzero expectation
value of a dimerized valence bond solid (VBS) order: $v \sim
(-1)^j \vec{S}_j \cdot \vec{S}_{j+1}$. As a side-note, we
emphasize that the state we are studying here is different from
the $\SO(3)$ or $\SU(2)$ SPT state defined through the group
cohomology of $\SO(3)$ or $\SU(2)$~\cite{wenspt,wenspt2,liuwen},
since in those states the symmetry acts chirally, $i.e.$ it only
acts on either the left or right modes. While in our case the spin
symmetry acts on both the left and right modes of the $1d$
boundary, and another discrete symmetry such as translation is
demanded.

Our goal is to study the edge states when the bulk undergoes a
disorder-order quantum phase transition, and the disordered phase
of the bulk phase diagram is the AKLT state. The quantum critical
fluctuation in the bulk may affect the edge of the AKLT state. To
study the interplay between the topologically protected edge
states, and the quantum critical modes, we adopt the ``two layer"
picture used in Ref.~\cite{boundary2020}: in layer-1, the system
remains a gapped AKLT state in the bulk with solid edge states
described by Eq.~\ref{H} and Eq.~\ref{H0}; in layer-2 the system
undergoes a phase transition between an ordinary {\it trivial}
disordered phase and an ordered phase. These two systems are glued
together at the boundary. We have used the common wisdom that the
transition between the SPT phase and the ordered phase is
generically in the same universality class as the transition
between an ordinary disordered phase and an ordered
phase~\footnote{This statement can be inferred based on the
observation that, the topological effects of many of the SPT
states can be captured by a nonlinear Sigma model plus a
topological $\Theta-$term at $\Theta =
2\pi$~\cite{senthilashvin,xu3dspt}. The $\Theta=2\pi$ topological
term reduces precisely to a boundary term, and we do not expect
this topological term to change the bulk universality class.}. We
will discuss two kinds of ordered phases: an $\SO(3)$
antiferromagnetic order, and an Ising-like VBS order that
spontaneously breaks $T_x$, assuming the boundary is at $y = 0$.
In the bulk the two disorder-order transitions under discussion
correspond to the three dimensional ($3D$) $\SO(3)$ and Ising
Wilson-Fisher transitions respectively, which can be studied
through a standard $\epsilon = 4 - D$ expansion, where $D = 2+1$
is the space-time dimension in the bulk. We only extend the bulk
dimensionality of layer-2 to $3 - \epsilon$ spatial dimensions,
while the layer-1 still has a two-dimensional bulk and
one-dimensional boundary.

We denote the bulk $\SO(3)$ antiferromagnetic order parameter, and
the Ising-VBS order parameter in layer-2 as $\vec{\phi}$ and
$\phi$ respectively, which should couple to the N\'{e}el order
parameter $\vec{n}$ and the VBS order parameter $v$ at the
boundary theory of layer-1, and this coupling could lead to new
physics in the infrared. However, $\vec{\phi}$ and $\phi$ do not
directly couple to $\vec{n}$ and $v$ due to the boundary condition
of the Wilson-Fisher fixed point. Assuming the boundary of the
$2d$ system is at $y = 0$, the most natural boundary condition for
fields $\vec{\phi}, \phi$ would be $\vec{\phi}(y = 0) = \phi(y =
0) = 0$~\footnote{This boundary condition corresponds to the
``ordinary transition" in the standard boundary criticality
literatures; other possibilities can also occur such as special
and extraordinary boundary transitions~\cite{cardybook}.}. Then
the leading nonvanishing boundary fields with the same quantum
number as $\vec{\phi}$ and $\phi$ are $\vec{\Phi} \sim
\partial_y \vec{\phi}$ and $\Phi \sim
\partial_y \phi$~\cite{cardybook}.

The $\SO(3)$ order parameter $\vec{\phi}$ and the Ising order
parameter $\phi$ will not become critical simultaneously without
fine-tuning, but they can be treated in the same framework. The
boundary quantum critical modes $\vec{\Phi}$ and $\Phi$ couple to
the fields at the boundary of layer-1 through the following terms
in the action \beqn \mathcal{S} &=& \int d^2 \mathbf{x} \ g_n
\vec{\Phi}(\mathbf{x}) \cdot \vec{n} (\mathbf{x}) + g_v \Phi
(\mathbf{x}) v(\mathbf{x}) \cr\cr &+& \int d^2\mathbf{x}
d^2\mathbf{x}' \ \frac{1}{2} \Phi^a (\mathbf{x})
C^{-1}_n(\mathbf{x}, \mathbf{x}')_{ab} \Phi^b(\mathbf{x}') \cr\cr
&+& \int d^2\mathbf{x} d^2\mathbf{x}' \ \frac{1}{2} \Phi
(\mathbf{x}) C^{-1}_v(\mathbf{x}, \mathbf{x}') \Phi(\mathbf{x}'),
\label{H2} \eeqn where $\mathbf{x} = (x, \tau)$ is the space-time
coordinate. $C_n(\mathbf{x}, \mathbf{x}')_{ab}$ and
$C_v(\mathbf{x}, \mathbf{x}')$ are the normalized correlation
functions of $\Phi^a$ and $\Phi$ at the boundary: \beqn &&
C_n(\mathbf{x}, 0)_{ab} = \langle \Phi^a (x, \tau)
\Phi^b(0,0)\rangle = \frac{\delta_{ab}}{(x^2 + \tau^2)^{3/2 -
\epsilon_n}}, \cr\cr && C_v(\mathbf{x}, 0) = \langle \Phi(x, \tau)
\Phi(0,0)\rangle = \frac{1}{(x^2 + \tau^2)^{3/2 - \epsilon_v}}.
\eeqn The scaling dimension of $\vec{\Phi}$ and $\Phi$ is
$\Delta_n = D/2 - \epsilon_n + O(\epsilon^2)$ and $\Delta_v = D/2
- \epsilon_v + O(\epsilon^2)$, where $D = 3$ is the bulk
space-time dimension. $\epsilon_{n/v}$ can be computed again
through the $\epsilon = (4 - D)$ expansion, following the
calculation of boundary criticality of the Wilson-Fisher fixed
points~\cite{cardybook,boundary2,boundary3,boundary4,boundary5}:
for an O($N$) Wilson-Fisher fixed point in the bulk, the scaling
dimension of the boundary modes of the order parameter is \beqn
\Delta_{\O(N)} = \frac{D}{2} - \frac{N + 2}{2(N + 8)}\epsilon +
O(\epsilon^2). \eeqn In our case $\epsilon_{n/v} = \epsilon (N +
2)/(2(N + 8)) $ with $N = 3, 1$ respectively. We again stress that
the $\epsilon$ dimensionality was introduced for layer-2 only. The
effective action of $\vec{\Phi}$ and $\Phi$ in Eq.~\ref{H2}
already received leading order correction from the
$\epsilon-$expansion due to the self-interaction of the bulk
critical modes. These effective actions can in principle receive
further corrections from the $g_v$ and $g_n$ couplings with the
boundary fields $\vec{n}$ and $v$, but this correction should be
at least at the order of $g_n^2, g_v^2$, which will be at higher
order of $\epsilon-$expansion. As we can see later, the main
physics we will discuss is at the vicinity of a fixed point where
$g_n, g_v \sim \epsilon$.

Eq.~\ref{H0}, \ref{H1}, \ref{H2} together can be viewed as an
effective non-local $1d$ theory, and this theory will be the
starting point of our discussion hereafter. Considering the fact
that the scaling dimension of both the N\'{e}el and VBS order
parameter at the $\SU(2)_1$ CFT is $1/2$, to the leading order of
$\epsilon$ expansion, the scaling dimensions of the coupling
constants must be \beqn && \Delta_{g_n} = \epsilon_n +
O(\epsilon^2), \ \ \Delta_{g_v} = \epsilon_v + O(\epsilon^2)
\cr\cr && \epsilon_n = \frac{5}{22}\epsilon, \ \ \ \epsilon_v =
\frac{1}{6}\epsilon. \eeqn $g_{n/v}$ are hence weakly relevant
assuming a small parameter $\epsilon$. Hence the $\SU(2)_1$ CFT at
the boundary of the AKLT state will be unstable against coupling
to the quantum critical modes, while fortunately due to the weak
relevance of the coupling constants, this effect can be studied
perturbatively.

\begin{figure}
\includegraphics[width=0.45\textwidth]{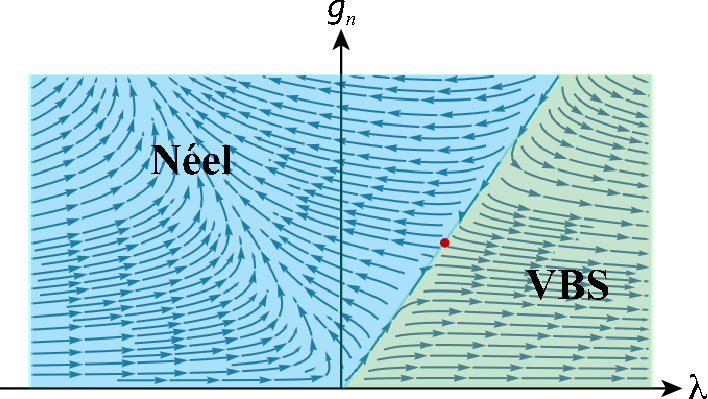}
\caption{The coupled RG flow of $\lambda$ and $g_n$ based on
Eq.~\ref{rg1}. A new fixed point $(\lambda^\ast, g^\ast_n) =
(\frac{2 \epsilon_n}{\pi}, \ \frac{4 \epsilon_n}{\pi}$) is found,
which separates two phases: the phase where $\lambda \rightarrow +
\infty$ is the VBS phase, and the phase with $(\lambda, g_n)
\rightarrow (- \infty, + \infty)$ is the long range N\'{e}el order
at the $1d$ boundary. But on the N\'{e}el order side of the phase
diagram, the RG flow is complicated and nonmonotonic, hence it may
take a long RG scale, or a large system size to finally reveal the
true long range order.} \label{fig:RG_flow}
\end{figure}

To proceed we need to compute the coupled renormalization group
(RG) flow of $\lambda$ and $g_{n/v}$ in Eq.~\ref{H1} and
Eq.~\ref{H2}. The RG equations can be derived based on the
following operator product expansion (OPE): \beqn && J^a_L(z)
n^b(w, \bar{w}) \sim \frac{1}{2}\frac{1}{z - w} \left( \ii
\delta_{ab} v(w, \bar{w}) + \ii \epsilon_{abc} n^c(w,\bar{w})
\right), \cr\cr && J^a_R(\bar{z}) n^b(w, \bar{w}) \sim
\frac{1}{2}\frac{1}{\bar{z} - \bar{w}} \left( - \ii \delta_{ab}
v(w, \bar{w}) + \ii \epsilon_{abc} n^c(w,\bar{w}) \right), \cr\cr
&& J^a_L(z) v(w, \bar{w}) \sim - \frac{1}{2} \frac{\ii }{z - w}
n^a(w,\bar{w}), \cr\cr && J^a_R(\bar{z}) v(w, \bar{w}) \sim
\frac{1}{2} \frac{\ii}{\bar{z} - \bar{w}} n^a(w,\bar{w}).
\cr\cr\cr && \left( \sum_a \nop{z}{a} \Phi^a(z, \bar{z}) \right)
\left( \sum_b \nop{w}{b} \Phi^b(w, \bar{w}) \right) \cr\cr && \sim
\frac{3}{2} \frac{1}{|z-w|^4} +\frac{1}{2} \frac{1}{|z-w|^2}
\sum_{a=1,2,3} J^a_L(w) J^a_R(\bar{w}), \cr\cr && ~~~+ \frac{3}{4}
\frac{1}{(\bar{z}-\bar{w})^{2 } } T_L(w) + \frac{3}{4}
\frac{1}{(z-w)^{2} } T_R(\bar{w}) +..., \cr\cr\cr && \left(
\vop{z} \Phi(z, \bar{z}) \right) \left( \vop{w} \Phi(w, \bar{w})
\right) \cr\cr && \sim \frac{1}{2} \frac{1}{|z-w|^4} - \frac{1}{2}
\frac{1}{|z-w|^2} \sum_{a=1,2,3} J^a_L(w) J^a_R(\bar{w}) \cr\cr &&
~~~+ \frac{1}{4} \frac{1}{(\bar{z}-\bar{w})^{2 } } T_L(w) +
\frac{1}{4} \frac{1}{(z-w)^{2} } T_R(\bar{w}) +...,
 \cr\cr\cr && \left( \sum_{a=1,2,3} J^a_L(z)
J^a_R(\bar{z}) \right) \left( \sum_{b=1,2,3} J^b_L(w)
J^b_R(\bar{w}) \right) \cr\cr && \sim \frac{3}{4}
\frac{1}{|z-w|^4} - \frac{2}{|z-w|^2} \sum_{a=1,2,3} J^a_L(w)
J^a_R(\bar{w}) \cr\cr && ~~~+ \frac{3}{2}
\frac{1}{(\bar{z}-\bar{w})^{2 } } T_L(w) + \frac{3}{2}
\frac{1}{(z-w)^{2} } T_R(\bar{w}) +... \label{ope} \eeqn In these
equations, $z$ and $w$ are the chiral coordinates ($z = \tau + \ii
x$); and the ellipsis contains less singular terms of the OPEs.
The fields $T_{L/R}$ are the energy-momentum tensor of the left
and right movers, which are given via the Suguwara construction by
$T_{L} = \frac{1}{3}\sum_a :J^a_L J^a_L:$ and $T_{R} = \frac{1}{3}
\sum_a :J^a_R J^a_R:$. Notice the form of energy-momentum tensors
is similar to the Hamiltonian Eq. \ref{H0} but with an extra
factor of $2\pi$. The OPEs above involving the fields $\Phi^a$ and
$\Phi$ are derived to the leading order of $\epsilon_{n/v}$.

These OPEs are sufficient to derive the desired RG equations to
the second order of the coupling constants. For example, using the
first two lines of Eq.~\ref{ope}, we can derive another set of
secondary OPEs: \beqn && \left( \sum_{a=1,2,3} J^a_L(z)
J^a_R(\bar{z}) \right) \left( \sum_b \nop{w}{b} \Phi^b(w, \bar{w})
\right) \cr\cr && \sim \frac{1}{4} \frac{1}{|z-w|^2} \left( \sum_b
\nop{w}{b} \Phi^b(w, \bar{w}) \right), \cr\cr\cr && \left(
\sum_{a=1,2,3} J^a_L(z) J^a_R(\bar{z}) \right) \left( \vop{w}
\Phi(w, \bar{w}) \right) \cr\cr && \sim -\frac{3}{4}
\frac{1}{|z-w|^2} \left( \vop{w} \Phi(w, \bar{w}) \right).
\label{ope2}\eeqn The coupled RG equations (beta functions) for
$\lambda$ and $g_{n/v}$ then read \beqn \beta(\lambda) =
\frac{d\lambda}{d\ln l} &=& 2 \pi \lambda^2 - \frac{\pi}{2} g_n^2
+ \frac{\pi}{2}g_v^2, \cr\cr \beta(g_n) = \frac{dg_n}{d\ln l} &=&
\epsilon_n g_n - \frac{\pi}{2} \lambda g_n, \cr\cr \beta(g_v) =
\frac{dg_v}{d\ln l} &=& \epsilon_v g_v + \frac{3\pi}{2} \lambda
g_v. \label{rg1} \eeqn These RG equations are valid as long as we
restrict our analysis to the parameter region with $\lambda, g_n,
g_v \sim \epsilon$, since every term in the RG equations
Eq.~\ref{rg1} would be at the same order of $\epsilon^2$.

As we explained before, there is no general reason for
$\vec{\phi}, \phi$ to become critical simultaneously in the bulk.
Hence let us ignore the $\Phi$ field first, and consider the
coupled RG equation for $\lambda, g_n$ only. If there is no bulk
quantum critical modes, an initial positive value $\lambda =
\lambda_0$ will be marginally relevant, and open up an energy gap
when it flows to positive infinite. According to the LSM theorem,
and the nature of the SPT state, this $1d$ boundary cannot be
trivially gapped, hence a nonperturbative positive $\lambda$ would
drive the system into an $\SO(3)$ invariant VBS state with
spontaneous symmetry breaking of translation symmetry $T_x$. But
by coupling to the boundary modes $\vec{\Phi}$ of quantum critical
fluctuation, the beta functions have an new unstable fixed point
at \beqn (\lambda^\ast, g^\ast_n) = \left(\frac{2
\epsilon_n}{\pi}, \ \frac{4 \epsilon_n}{\pi} \right). \eeqn The
two eigenvectors of RG flow expanded at the new fixed point have
scaling dimensions $(8.9 \epsilon_n, -0.89 \epsilon_n)$.

Of course the RG analysis above is only at the leading nontrivial
order of $\epsilon-$expansion, and at this order of accuracy, no
other fixed point is found in the phase diagram. The new fixed
point found above separates two phases: phase I where $\lambda$
flows to positive infinity, and phase II where $\lambda$ and $g_n$
flow to negative and positive infinity respectively. Then both
phases no longer have scaling invariance, so both phases should
have certain long range order considering the fact that there is
no topological order in one dimension~\cite{wenreview2}. Phase I
with $\lambda \rightarrow + \infty$ is the dimerized VBS phase as
we discussed before; phase II with $(\lambda, g_n) \rightarrow ( -
\infty, + \infty)$ should be a N\'{e}el ordered phase, $i.e.$ the
$1d$ boundary can develop the N\'{e}el order before the bulk, even
though the bulk is still at a quantum critical point. A negative
$\lambda$ would enhance the correlation of the N\'{e}el order
parameter, and after integrating out $\vec{\Phi}$, a long range
interaction proportional $g^2$ would be generated between the
N\'{e}el order parameters. Hence the infrared limits $\lambda
\rightarrow -\infty$ and $g \rightarrow + \infty$ of phase II both
favor the long range N\'{e}el order.

The correlation length critical exponent $\nu$ of this
N\'{e}el-VBS transition is $\nu \sim 1/(8.9 \epsilon_n)$. At the
transition point $(\lambda^\ast, g^\ast_n) = (2 \epsilon_n / \pi,
4 \epsilon_n/\pi)$, the scaling dimensions of the N\'{e}el and VBS
order parameters can again be computed to the leading order of
$\epsilon-$expansion: \beqn \Delta_{\vec{n}} &=& \frac{1}{2} +
\frac{\pi \lambda^\ast}{2} = \frac{1}{2} + \epsilon_n, \cr\cr
\Delta_{v} &=& \frac{1}{2} - \frac{3\pi \lambda^\ast}{2} =
\frac{1}{2} - 3\epsilon_n. \label{scaling}\eeqn One can see that
compared with the $\SU(2)_1$ CFT, the N\'{e}el order correlation
is suppressed while the VBS order correlation is enhanced at the
new transition fixed point, since $\lambda^\ast > 0$. This also
implies that this N\'{e}el-VBS transition has no enlarged symmetry
of $\SU(2)_L \times \SU(2)_R$. An enlarged $\SU(2)_L \times
\SU(2)_R \sim \SO(4)$ symmetry would guarantee that the N\'{e}el
and VBS order parameters have the same scaling dimension, because
$(\vec{n}, v)$ transform as a vector under $\SO(4)$. Many previous
studies suggest that at an unconventional quantum critical point
between two phases with different spontaneous symmetry breaking,
an enlarged emergent symmetry in the infrared is often expected
due to a series of
dualities~\cite{xudual,seiberg2,mrossdual,deconfinedual,potterdual,dualreview,xutriangle}.
But in our current case we expect the infrared symmetry at the
N\'{e}el-VBS transition is still the microscopic symmetry $\SO(3)
\times G$.

As we mentioned before, suppose we integrate out the field
$\vec{\Phi}$ in Eq.~\ref{H2}, a long range interaction in
space-time will be generated between the N\'{e}el order parameter.
The scenario is similar to the spin-1/2 chain with a long range
spin-spin interaction, the only difference is that in the latter
case the long range interaction is instantaneous and only nonlocal
in space. Recently a direct transition between the N\'{e}el and
VBS order was found in a spin-1/2 chain with nonlocal two-spin
interaction and local four-spin
interaction~\cite{sandvik1,sandvik2}. It was found numerically
that at the direct N\'{e}el-VBS transition the scaling dimension
of the N\'{e}el order parameter is greater than the VBS order
parameter, which is fundamentally different from the $\SU(2)_1$
CFT, but consistent with our RG calculations Eq.~\ref{scaling}. We
also note that a previous RG analysis was performed for $1d$
spin-1/2 system with an instantaneous nonlocal spin interaction,
but the N\'{e}el-VBS transition was not found therein. Instead the
previous analysis identified a transition between the true long
range N\'{e}el order and a quasi-long range order at the parameter
region $\epsilon_n < 0$ and $\lambda < 0$ with our
notation~\cite{rgnonlocal}.

So far we have assumed that the fields $\vec{n}, v$ and
$\vec{\Phi}, \Phi$ have the same velocity in our effective $1d$
theory Eq.~\ref{H2}, hence the theory we considered so far has a
Lorentz invariance. We can also turn on a weak velocity difference
between these two sets of fields, and analyze how it flows under
RG. This velocity anisotropy corresponds to modifying the
correlation function of $\vec{\Phi}$: \beqn C_n(\mathbf{x},
0)_{ab} = \langle \Phi^a (x, \tau) \Phi^b(0,0)\rangle \cr\cr =
\frac{\delta_{ab}}{\left( (1 - \frac{\delta v}{2})^2 x^2 + (1 +
\frac{\delta v}{2})^2 \tau^2 \right)^{3/2}}. \eeqn Here we have
assumed that the velocity of $\vec{\Phi}$ exceeds the velocity of
$\vec{n}$ by a factor of $(1 + \delta v)$ (to the first order of
$\delta v$). We have taken $\epsilon_n = 0$ for the leading order
calculation. $\delta v$ can flow under RG as it is the ``seed" for
velocity difference. Based on symmetry, the RG flow of $\delta v$
should look like \beqn \frac{d \delta v}{d\ln l} = - \alpha g^2_n
\delta v. \eeqn And eventually we will plug in the fixed point
value of $g_n = g_{n}^\ast$. Based on previous experience, at an
interacting fixed point, a weak velocity anisotropy is often
irrelevant~\cite{hermele2005,leesusy}, since intuitively in the
infrared all the interacting modes are expected to have the same
velocity. Hence we expect $\alpha > 0$, $i.e.$ a weak velocity
difference between the boundary and bulk will be irrelevant at the
N\'{e}el-VBS transition fixed point.

To evaluate $\alpha$, we expand the correlation function of
$\vec{\Phi}$ to the leading order of $\delta v$: \beqn
C_n(\mathbf{x}, 0) = \frac{1}{|z|^{3}} - \frac{3}{2} \frac{\delta
v}{|z|^5} \frac{z^2 + \bar{z}^2}{2} + O(\delta v^2) \eeqn Using
the OPEs in Eq.~\ref{ope2}, the second order perturabtion of $g_n$
would generate the following term: \beqn && - \frac{1}{2} \ g^2_n
\left( \sum_a \nop{z}{a} \Phi^a(z, \bar{z}) \right) \left( \sum_b
\nop{w}{b} \Phi^b(w, \bar{w}) \right) \cr\cr && \sim -
\frac{3g_n^2}{4|z-w|^4} - g_n^2 \frac{1}{4} \frac{1}{|z-w|^2}
\sum_{a=1,2,3} J^a_L(w) J^a_R(\bar{w}) \cr\cr && + \ g_n^2 \delta
v \frac{9}{32} \frac{1}{|z - w|^2 } \left( T_L(w) + T_R(\bar{w})
\right) + \cdots \label{velocity}\eeqn Here we only kept the terms
that will lead to nonzero effect under real space RG. The last
term in Eq.~\ref{velocity} would contribute a renormalization (or
acceleration) for the velocity of $\vec{n}$. Under rescaling, the
ratio between the two velocities reduces by a factor: \beqn 1 +
\delta v \rightarrow \frac{1 + \delta v}{ 1 + g_n^2 \delta v
\frac{9\pi^2}{8} \ln l}, \eeqn which leads to the RG equation for
$\delta v$: \beqn \frac{d \delta v}{ d\ln l} = -
\frac{9\pi^2}{8}(g_n^\ast)^2 \delta v, \eeqn which confirms our
expectation that $\delta v$ is an irrelevant perturbation at the
N\'{e}el-VBS transition fixed point.

Suppose we start with $\delta v > 0$, namely the velocity of
$\vec{n}$ is smaller than $\vec{\Phi}$, the velocity of $\vec{n}$
will increase under RG. This means that in this case the system
will qualitatively behave like $z < 1$, where $z$ is the dynamic
critical exponent (not to confuse with the chiral coordinate). On
the contrary, if we start with $\delta v < 0$, the velocity of
$\vec{n}$ would decrease under RG, which means that effectively $z
> 1$. The former scenario is analogous to a spin chain with
instantaneous spatial nonlocal interaction~\cite{sandvik2}, which
is equivalent to taking the velocity of the effective action of
$\vec{\Phi}$ and $\Phi$ to infinity in our effective $1d$ theory
Eq.~\ref{H2}. Although our calculation is for $\delta v > 0$,
rather than taking the velocity in the $\vec{\Phi}$ action to be
infinity, the ``acceleration" of the modes derived here (including
$z < 0$) is qualitatively consistent with what was observed in
Ref.~\cite{sandvik2} at the N\'{e}el-VBS transition in a spin-1/2
chain with nonlocal spatial interactions.

\begin{figure}[h]
\includegraphics[width=0.4\textwidth]{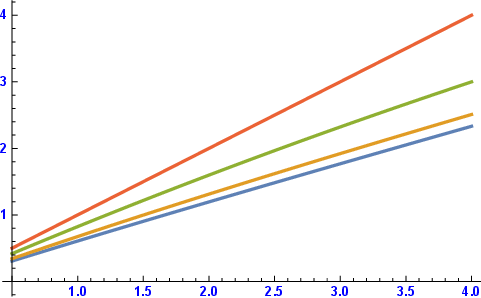}
\caption{ The plot of $\ln[ 3\pi G_n(\mathbf{k})(1 + A (g^{\ast
\prime}_n)^2)]$ against $\ln[1/|\mathbf{k}|]$, where
$G_n(\mathbf{k})$ is given by Eq.~\ref{corre}. From top to bottom,
$A (g^{\ast \prime}_n)^2 = 0, 1/2, 2$, and $5$.} \label{crossover}
\end{figure}

\begin{figure}[h]
\includegraphics[width=0.45\textwidth]{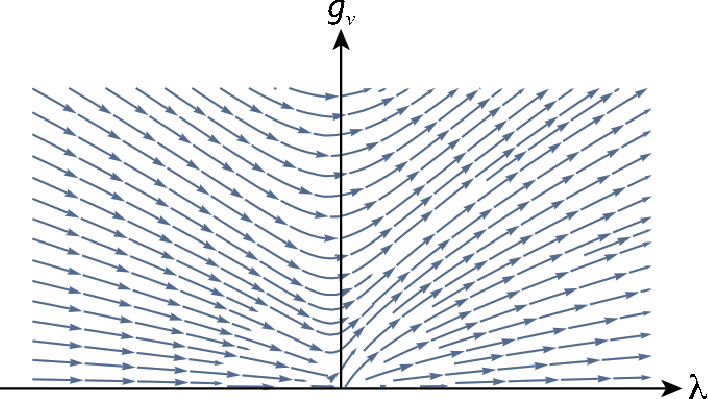}
\caption{The RG flow of $(\lambda, g_v)$. As long as the initial
value $g_v$ is nonzero, both parameters will flow to positive
infinity, which implies that the boundary will likely develop the
Ising-VBS order before the bulk. } \label{RG02}
\end{figure}

In the phase diagram Fig.~\ref{fig:RG_flow}, on the side of the
N\'{e}el order, the path of the RG flow towards the long range
order can be complicated. It may take a long RG scale and hence
large system size to reveal the true long range order. For
example, on part of the phase diagram, $\lambda$ changes its sign
and eventually flow away to the negative nonperturbative regime.
While $\lambda$ changes sign, $g_n$ first decreases its magnitude
from the initial value $g_0$, then after reaching its minimum
$g^{\ast \prime}_n$ along the RG flow, $g_n$ keeps increasing and
eventually become nonperturbative. Hence it is possible that for a
relatively large intermediate scale, the system behaves like $g_n
\sim g^{\ast \prime}_n$. The effect of this nonmonotonic RG flow
can be illustrated by a simple perturbation theory to the
correlation function of the N\'{e}el order parameter: \beqn &&
G_n(\mathbf{x}) = \langle \vec{n}(\mathbf{x}) \cdot \vec{n}(0)
\rangle  \cr\cr &\sim& \frac{3}{2}\frac{1}{|\mathbf{x}|} +
\frac{3}{4} \int d^2\mathbf{x}_1 d^2\mathbf{x}_2 \frac{(g^{\ast
\prime}_n)^2}{|\mathbf{x} - \mathbf{x}_1||\mathbf{x}_1 -
\mathbf{x}_2|^{3 - 2 \epsilon_n}|\mathbf{x}_2|} \cr\cr &+&
O(g^{\ast \prime}_n)^4 + \cdots. \label{corre}\eeqn Hence
$G_n(\mathbf{k})$ in the momentum-frequency space $\mathbf{k} =
(k, \omega)$ reads \beqn G_n(\mathbf{k}) \sim
\frac{1}{G^{(0)}(\mathbf{k})^{-1} - \Sigma(\mathbf{k})}, \eeqn
where $G^{(0)}(\mathbf{k}) = 3\pi / |\mathbf{k}| $,
$\Sigma(\mathbf{k}) = - A (g^{\ast \prime}_n)^2 |\mathbf{k}|^{1 -
2 \epsilon_n} / (3\pi) $, and $A > 0$ for $0 < \epsilon_n < 1/2$.
The system will have enhanced spin-spin correlation function
compared with the $\SU(2)_1$ CFT of the spin-1/2 chain, as was
observed in numerical simulations~\cite{zhang1,stefan1,stefan2}.
The mixture of the two terms in $G^{-1}(\mathbf{k})$ may yield
results that appear to be power-law correlation with different
scaling dimensions, which is illustrated in Fig.~\ref{crossover},
where we have fixed $\epsilon_n = 5/22 \epsilon$ but chosen
different $g^{\ast \prime}_n$. This nonuniversal power-law like
scaling of spin correlation was also observed in recent numerics
concerning the edge states of the AKLT state during a bulk phase
transition~\cite{stefan1,stefan2}.

Now we briefly consider the situation when the bulk undergoes a
disorder-order quantum phase transition between the AKLT state and
the Ising like VBS order, which is described by order parameter
$\phi$. The boundary mode of $\phi$ is $\Phi \sim \partial_y
\phi$, and it couples to the VBS order parameter $v$ at the
boundary CFT. In this case, the coupled RG flow of $\lambda$ and
$g_v$ in Eq.~\ref{H2} is relatively simple: as long as we start
with nonzero $(\lambda_0, g_{v0})$, both $g_v$ and $\lambda$ quite
generally flow to positive infinity, which corresponds to a
nonzero long range order of $v$. Hence the $1d$ boundary of the
system should develop the Ising-VBS order before the bulk. when
the bulk is tuned closer and closer to a VBS (Ising) transition,
the boundary will go through a transition between the gapless
$\SU(2)_1$ CFT state to a VBS phase, before the bulk actually hits
criticality. This boundary transition should be in the same
universality class as the transition from an $\SU(2)_1$ CFT to a
VBS phase in a purely one-dimensional spin-1/2 chain with both
nearest and next nearest neighbor Heisenberg interactions (see,
for example, Ref.~\onlinecite{Gogolin2004} for the one-dimensional
transition). We note that this transition is not an ordinary
$1+1d$ Ising transition and, hence, is different from the
``extraordinary transition" studied in the standard boundary
criticality literature. But if we start with a negative initial
value $\lambda_0$, it may take a long RG time before the coupling
constants become positive and nonperturbative. Hence the VBS order
parameter may still appear to have quasi long range correlation
for a finite system.

In conclusion, we have found that there can be a direct continuous
quantum phase transition between the long range antiferromagnetic
N\'{e}el order, and the VBS order, in an effective $1d$ spin-1/2
system with nonlocal interactions (Eq.~\ref{H2}). Due to the
nonlocality of the model, even in a $1d$ system with a continuous
$\SO(3)$ spin symmetry there can be a long range N\'{e}el order.
Within the accuracy of our method, the effective spin-1/2 system
Eq.~\ref{H2} arises from coupling the $1d$ boundary of a $2d$ SPT
phase to bulk quantum critical modes. Our results were drawn from
a controlled renormalization group study, and the critical
exponents extracted (including the anomalous dimensions of order
parameters and the dynamical exponent) are qualitatively
consistent with the N\'{e}el-VBS transition found numerically in
recent simulation of a spin-1/2 chain with spatially instantaneous
nonlocal interactions~\cite{sandvik1,sandvik2}. If a $1d$ system
has local interactions only, there can only be spontaneous
discrete symmetry breaking. Previous numerical and analytical
works~\cite{sandvik3,motrunich1d1,motrunich1d2} have studied the
analogue of deconfined quantum critical point between two phases
that spontaneously break different discrete symmetries.

This work is supported by NSF Grant No. DMR-1920434, the David and
Lucile Packard Foundation, and the Simons Foundation. The authors
thank Anders Sandvik and Leon Balents for helpful discussions.

\bibliography{1db}

\end{document}